\documentclass[11pt]{article}

\usepackage[preprint]{acl}

\usepackage{times}
\usepackage{latexsym}
\usepackage{booktabs}   
\usepackage{multirow}   
\usepackage{amsmath}    
\usepackage{amssymb}   

\usepackage[T1]{fontenc}

\usepackage[utf8]{inputenc}
\usepackage{graphicx}
\usepackage{subcaption}
\usepackage{microtype}

\usepackage{inconsolata}

\title{Dependency-Guided Code Generation: Structured Matrix Decomposition and Consistency-Guided Refinement}

\author{
  \textbf{Mingqiao Mo}$^{1}$,
  \textbf{Yangchen Zeng}$^{2}$,
  \textbf{Zikai Xiao}$^{3}$,
  \textbf{Xin Xiao}$^{1}$,
  \textbf{Wenhua Nie}$^{4}$ \\
  \textbf{Zhaolu Kang}$^{5}$,
  \textbf{Guangyuan Dong}$^{6}$,
  \textbf{Kai Shu}$^{7}$,
  \setcounter{footnote}{1}%
  \textbf{Hao Zhang}$^{1}$\thanks{Corresponding authors.},
  \textbf{Xiaodong Fan}$^{8}$\footnotemark[2] \\
  \\
  $^{1}$University of the Chinese Academy of Sciences,
  $^{2}$Southeast University,
  $^{3}$Zhejiang University \\
  $^{4}$National Taiwan University,
  $^{5}$Peking University,
  $^{6}$National University of Singapore \\
  $^{7}$Tsinghua University,
  $^{8}$Liaoning Technical University
}

\begin{document}
\maketitle
\begin{abstract}
The increasing complexity of modern software systems has made automated code generation a fundamental task in software engineering. However, existing approaches often fail to adequately capture the intricate, multi-level dependencies among code entities, leading to generated code that is logically incomplete or difficult to integrate into real-world systems. To address this limitation, we propose a dependency-aware code generation framework that explicitly models interactions among code entities through a graph-based representation. We decompose dependencies into two complementary components: a quantized matrix that captures strong, explicit relations, and a sparse low-rank factorization that models weaker, implicit interactions. The decomposition is efficiently learned via an alternating optimization procedure. During code generation, the learned dependency structure is incorporated as a constraint, ensuring both semantic coherence and structural consistency of the generated code. Furthermore, we introduce a sparse triplet representation for strong dependencies, significantly improving storage efficiency and computational scalability. Extensive experiments demonstrate that our approach consistently produces code with superior semantic alignment and structural fidelity compared to existing methods.
\end{abstract}

\section{Introduction}
\label{sec:intro}

The rapid evolution of modern software systems has significantly increased the complexity and scale of codebases, making automated code generation a fundamental problem in software engineering~\cite{LeGoues2021, Weimer2009, Cadar2013, Chen2021}. Recent advances in large-scale pretraining and the availability of massive open-source repositories have enabled models to automatically synthesize functions, modules, and even end-to-end programs with impressive fluency~\cite{li2022competition,chen2022codet,svyatkovskiy2020intellicode}. The broader impact of large language models (LLMs) extends well beyond code, spanning mathematical reasoning~\cite{mo2026pathsymphony}, intent understanding under ambiguity~\cite{he2025enhancing}, multimodal reasoning~\cite{kang2026multimodal}, and even security analysis~\cite{wu2025sugar}, underscoring their versatility but also their susceptibility to subtle failures when structural constraints are not explicitly enforced. These developments hold the promise of reducing human effort, accelerating development cycles, and improving software maintainability~\cite{Motwani2023, Noller2022, Jiang2023,dehaerne2022code}. Moreover, learning robust code representations under diverse execution environments, such as virtual-machine-protected code, has attracted growing attention~\cite{mo2026shieldedcode}, highlighting the need for structural awareness in code modeling. 

Despite this progress, generating code that is not only syntactically correct but also \emph{structurally consistent} and \emph{system-aware} remains a major challenge. In real-world settings, code does not exist in isolation: functions depend on other functions, classes interact through inheritance and composition, and modules are tightly coupled through imports and shared interfaces. Existing generation approaches often fail to capture these interdependencies, resulting in outputs that are locally plausible but globally inconsistent, difficult to integrate, or prone to subtle bugs.

Existing code generation methods can be broadly categorized into three groups. 
\textbf{(1) Statistical and search-based methods} explore the program space via mutation, templates, or heuristic search, providing diversity but lacking mechanisms to enforce global structural coherence~\cite{Ren2022, Zheng2023}. 
\textbf{(2) Semantic and constraint-based approaches} leverage type systems, symbolic execution, or formal specifications to ensure correctness, but often struggle with scalability and incomplete specifications in large systems~\cite{Weimer2009, Nguyen2013, Mechtaev2016}. 
\textbf{(3) Deep learning and large language model (LLM) approaches} learn rich syntactic and semantic patterns from data and achieve strong performance in code completion and synthesis tasks, yet they primarily rely on local context or corpus-level co-occurrence statistics, without explicitly modeling system-level dependencies~\cite{Jimenez2023, Pu2023a}. Meanwhile, adaptive prompt optimization~\cite{zhang2026adaptive} and probability-entropy calibration~\cite{yu2026probability} have improved LLM alignment but still do not address the fundamental gap in structural dependency reasoning. 

A key limitation shared by these paradigms is the lack of an explicit representation of \emph{multi-level dependency structures} among code entities. In practice, software systems exhibit both strong, discrete dependencies (e.g., function calls, inheritance hierarchies) and weaker, implicit relationships (e.g., indirect coupling, shared design patterns). Ignoring this distinction leads to generated code that may violate critical constraints or fail to align with the architectural structure of the codebase.

\begin{figure*}[t]
  \centering
  \includegraphics[width=1\linewidth]{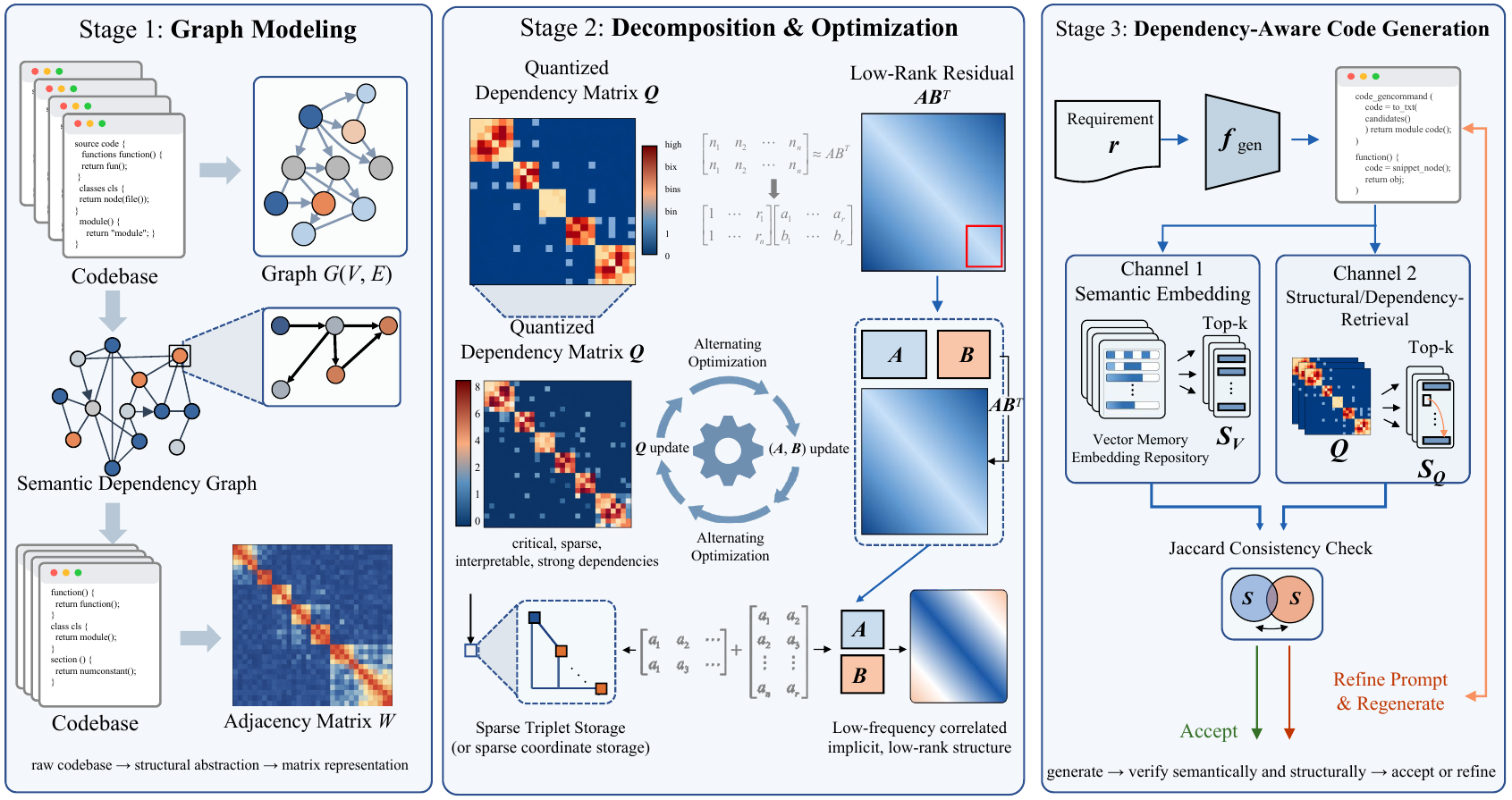}
  \caption{\textbf{Overview of the proposed dependency-aware code generation framework.}
  \label{111}
\textit{Left:} We first construct a semantic dependency graph from the input codebase, where nodes represent code entities (e.g., functions, classes, files) and edges encode their interactions, which are further summarized as an adjacency matrix capturing inter-entity relationships.
\textit{Middle:} The dependency matrix is decomposed into two complementary components: a quantized matrix ($Q$) that models strong, discrete dependencies, and a low-rank residual ($AB^\top$) that captures weak and implicit interactions. These components are jointly optimized via an alternating minimization procedure, with sparse triplet storage enabling scalability.
\textit{Right:} During generation, the learned dependency structure is integrated into a dual-channel process combining semantic retrieval and structural dependency matching. A consistency check (e.g., Jaccard similarity) enforces alignment between the two views, and the model iteratively refines the generated code until both semantic coherence and structural consistency are satisfied.}
\end{figure*}

To address this gap, we propose a \emph{dependency-aware code generation framework} that explicitly models and enforces relationships among code entities. We represent a codebase as a graph, where nodes correspond to entities such as functions, classes, and files, and edges encode their interactions. Building on this representation, we introduce a hybrid decomposition of the dependency matrix into two complementary components: 
(i) a \textbf{quantized matrix} that captures strong, discrete, and interpretable dependencies, and 
(ii) a \textbf{sparse low-rank factorization} that models weak, implicit, and higher-order interactions. 
This formulation enables a principled separation between critical and non-critical dependencies, improving both interpretability and robustness.

We develop an efficient alternating optimization algorithm to jointly learn these components, and further design a sparsity-aware storage scheme that scales to large codebases. Crucially, the learned dependency structure is integrated into the code generation process: candidate outputs are evaluated against both semantic similarity and structural consistency, and iteratively refined until they satisfy dependency constraints. This mechanism ensures that generated code aligns with the underlying architecture of the system, leading to improved usability and reduced integration errors. Figure \ref{111} illustrates an overview of our method.
Our main contributions are summarized as follows:
\begin{itemize}
    \item We propose a dependency-aware code generation framework that models code entities and their interactions via a graph-based representation, and integrates dependency constraints into the generation process through a consistency-driven refinement mechanism.
    
    \item We introduce a hybrid dependency decomposition that combines a quantized matrix for strong, explicit dependencies with a sparse low-rank component for weak, implicit interactions, optimized via an efficient alternating scheme.
    
    \item We design a sparse triplet representation for strong dependencies, enabling scalable storage and efficient computation for large-scale codebases.
\end{itemize}

\section{Related Work}

\subsection{Pretrained Models and Structural Injection}
Pretrained code language models and instruction-tuned systems have advanced code generation by producing syntactically valid and semantically meaningful programs. Recent work improves this via large-scale pretraining, instruction tuning, and structure-aware objectives. CodeT5+~\cite{wang2023codet5plus} uses multiple pretraining objectives, while Magicoder~\cite{wei2024magicoder} builds synthetic coding instructions from open-source repositories to capture diverse programming patterns. Beyond scaling and data, methods inject program structure into learning. AST-T5~\cite{gong2024astt5} incorporates abstract syntax trees via structure-aware objectives, and IRCoder~\cite{paul2024ircoder} uses compiler intermediate representations to improve multilingual generation and transfer. Parallel efforts in efficient fine-tuning, such as probability-entropy calibration~\cite{yu2026probability}, have reduced the cost of adapting large models to downstream code tasks. Token-level efficiency techniques such as adaptive visual token pruning~\cite{zhang2025trimtokenator} and prefill-decode pruning~\cite{zhang2026pdtrim} further improve inference scalability. While these approaches strengthen code modeling, they mainly operate on functions or local fragments and do not explicitly model repository-scale dependencies or key cross-file interactions.

\subsection{Repository-Level Code Generation}
Repository-level code generation addresses multi-file software projects requiring reasoning over APIs, dependencies, and long-range cross-file context. Recent benchmarks emphasize this setting. RepoBench~\cite{liu2024repobench} evaluates repository-level code completion via retrieval and end-to-end tasks, while SWE-bench~\cite{jimenez2024swebench} shows real-world issue resolution requires coordinated edits across multiple files. Prior work explores retrieval, graph construction, and agentic tools. CoCoMIC~\cite{ding2024cocomic} models in-file and cross-file context, DraCo~\cite{cheng2024draco} builds dataflow-based context graphs for retrieval, and GraphCoder~\cite{liu2024graphcoder} uses coarse-to-fine graph-based retrieval. CodeAgent~\cite{zhang2024codeagent} integrates external tools for coding and testing, while RLCoder~\cite{wang2025rlcoder} optimizes retrieval and generation via reinforcement learning. Graph-structured representations have also proven effective in broader settings, including text-attributed graph learning~\cite{zhang2025can,zhang2026mitigating} and visual geo-localization with heterogeneous graph networks~\cite{zheng2025graphgeo}, demonstrating the general utility of graph-based modeling for capturing complex relational structures. Despite these advances, most methods rely on repository graphs as retrieval or prompt scaffolds. In contrast, our framework learns dependency representations and injects them into generation via a consistency-driven refinement mechanism, enabling more structured repository-level generation.

\section{Methodology}\label{sec:method}

\subsection{Overview}
We propose a graph-based framework to explicitly model dependency structures among code entities, by decomposing the dependency matrix into two complementary components: a quantized matrix $Q$, which captures strong and discrete structural dependencies, and a low-rank component $AB^\top$, which models weak, implicit, and higher-order interactions. Specifically, $Q$ encodes stable program-level relations such as function calls, inheritance, and imports, which are sparse but highly reliable, while $AB^\top$ captures distributed latent interactions induced by indirect dependencies, shared contexts, and semantic coupling between code entities. This decomposition is motivated by the observation that real-world codebases exhibit both dominant explicit structure and diffuse latent interactions; modeling them jointly in a single representation often leads to either over-smoothing or loss of structural fidelity. By separating these two factors, our approach provides a more balanced and interpretable representation of software structure. The model is learned via alternating optimization over $Q$, $A$, and $B$, and is further supported by sparsity-aware storage for scalability, with the resulting dependency structure integrated into the generation process to enforce both semantic consistency and structural correctness.

\subsection{Code Entity Dependency Modeling}

We represent a codebase as a directed graph $G=(V,E)$, where nodes correspond to code entities such as functions, classes, or files, and edges represent dependency relations such as function calls, inheritance, and module imports. The graph is encoded as an adjacency matrix $W \in \mathbb{R}^{n \times n}$, where each entry $W_{ij}$ measures the interaction strength between entities $i$ and $j$.

In practice, $W$ contains heterogeneous signals: some entries correspond to explicit program-level dependencies, while others reflect weak or noisy correlations induced by shared usage patterns or indirect coupling. This mixture makes direct modeling difficult, especially for downstream generation tasks that require both precision and generalization.

To address this, we decompose $W$ into two interpretable components:
\begin{equation}
W \approx Q + AB^\top.
\end{equation}

Here, $Q$ serves as a discrete structural backbone of the codebase, while $AB^\top$ provides a continuous latent correction that captures residual dependencies not explained by explicit edges.

We learn this decomposition by minimizing the following objective:
\begin{equation}
\small
\min_{Q,A,B} \|W - Q - AB^\top\|_F^2 + \lambda_A \|A\|_F^2 + \lambda_B \|B\|_F^2.
\end{equation}

The regularization terms encourage stable low-rank factors and prevent overfitting to noisy dependencies.

To make $Q$ explicitly interpretable, we define a discrete codebook $\{q_1,\dots,q_N\}$, where each level represents a different strength of dependency. Each entry $Q_{ij}$ is assigned via a one-hot variable $Z_{ij}^k$ satisfying $\sum_k Z_{ij}^k = 1$, leading to:
\begin{equation}
Q_{ij} = \sum_{k=1}^N q_k Z_{ij}^k.
\end{equation}

This quantization enforces that structural dependencies are not only sparse but also semantically categorized into discrete levels of importance, improving interpretability and robustness.

\subsection{Dependency Matrix Estimation}

We optimize the model using an alternating minimization strategy that iteratively updates the quantized component $Q$ and the low-rank factors $(A,B)$. This separation allows each component to be optimized under a simplified subproblem.

\paragraph{Alternating updates.}
When $A$ and $B$ are fixed, we first compute the residual matrix:
\begin{equation}
R^Q = W - AB^\top.
\end{equation}

This residual removes latent effects and isolates strong structural signals. We then update $Q$ by projecting each entry of $R^Q$ onto the nearest level in the predefined codebook:
\begin{equation}
Q = \mathcal{P}(R^Q),
\end{equation}
where $\mathcal{P}(\cdot)$ denotes element-wise quantization. This step can be interpreted as a denoising operation that extracts stable, high-confidence dependencies while suppressing weak fluctuations.

When $Q$ is fixed, we compute:
\begin{equation}
R^{AB} = W - Q,
\end{equation}
and learn a low-rank approximation:
\begin{equation}
\min_{A,B} \|R^{AB} - AB^\top\|_F^2 + \lambda_A \|A\|_F^2 + \lambda_B \|B\|_F^2.
\end{equation}

This step captures global interaction structure, including indirect dependencies (e.g., A influences C through B) and shared latent functionality across modules. Optionally, $\ell_1$ regularization can be applied to encourage sparsity in $A$ and $B$, further improving interpretability.

Overall, this alternating scheme progressively separates discrete structural dependencies from continuous latent interactions, leading to a more stable and semantically meaningful representation.

\subsection{Code Generation with Dependency Awareness}

To incorporate dependency information into code generation, we maintain a vector-based repository where each code entity is embedded using a pretrained encoder. Given a requirement $r$, the model generates a candidate code snippet $\hat{c}$, which is embedded as $v_{\hat{c}}$.

We retrieve semantically similar entities using embedding similarity:
\begin{equation}
\mathcal{S}_v = \mathrm{Top}\text{-}k(\cos(v_{\hat{c}}, v_c)).
\end{equation}

This captures functionally or syntactically similar code components.

In parallel, we retrieve structurally related entities using the learned dependency matrix:
\begin{equation}
\mathcal{S}_Q = \mathrm{Top}\text{-}k(Q_{\hat{c},:}).
\end{equation}

This ensures that generated code respects the underlying program structure, such as required imports, dependencies, or inherited behaviors.

To enforce consistency between semantic and structural views, we compute a Jaccard similarity:
\begin{equation}
J = \frac{|\mathcal{S}_v \cap \mathcal{S}_Q|}{|\mathcal{S}_v \cup \mathcal{S}_Q|}.
\end{equation}

If $J \ge \tau$, the generated code is considered consistent and accepted. Otherwise, we refine generation by augmenting the prompt with structurally relevant entities from $\mathcal{S}_Q$, explicitly steering the model toward dependency-consistent outputs. This process is repeated until convergence.

\subsection{Sparse Storage of Strong Dependencies}

Although $Q$ captures all pairwise dependencies, most entries correspond to weak or negligible interactions. Storing all entries is therefore inefficient and unnecessary.

To improve efficiency, we retain only significant entries:
\begin{equation}
\mathcal{T}(Q) = \{(i,j,Q_{ij}) \mid Q_{ij} \ge q_k\}.
\end{equation}

This reduces storage complexity from $O(n^2)$ to $O(|\mathcal{T}(Q)|)$, where $|\mathcal{T}(Q)| \ll n^2$ in real-world codebases.

Importantly, this sparse representation preserves interpretability because each retained triplet corresponds to a strong and meaningful dependency between two code entities. The resulting structure can be efficiently implemented using CSR or CSC formats, enabling scalable computation while remaining compatible with downstream retrieval and generation tasks. This design is in spirit similar to broader efficiency-driven approaches such as prototype-based incremental learning~\cite{wu2026protoflow} and hierarchical multi-scale feature fusion~\cite{cong2025hierarchical}, where reducing redundancy without losing critical structural information is a central goal.

\begin{table*}[t]
\centering
\caption{Comprehensive performance comparison on code translation tasks, including Java$\rightarrow$C\# and C\#$\rightarrow$Java, where we report BLEU, Exact Match (xMatch), and CodeBLEU to evaluate lexical, structural, and semantic correctness of generated code.}
\label{tab:translation}
\small
\begin{tabular}{lccc|ccc}
\toprule
& \multicolumn{3}{c}{Java $\rightarrow$ C\#} 
& \multicolumn{3}{c}{C\# $\rightarrow$ Java} \\
Model & BLEU & xMatch & CodeBLEU & BLEU & xMatch & CodeBLEU \\
\midrule
Transformer & 56.1 & 33.2 & 63.9 & 50.6 & 38.1 & 61.8 \\
CodeBERT & 80.3 & 59.4 & 85.3 & 72.6 & 59.1 & 80.2 \\
GraphCodeBERT & 81.0 & 60.0 & 86.1 & 73.2 & 60.2 & 81.0 \\
PLBART & 83.5 & 65.1 & 88.0 & 78.9 & 65.6 & 85.9 \\
CodeT5 & 84.2 & 66.0 & 88.5 & 80.0 & 67.3 & 86.2 \\
\midrule
\textbf{Ours} & \textbf{86.1} & \textbf{69.4} & \textbf{90.3} 
& \textbf{82.2} & \textbf{70.1} & \textbf{88.1} \\
\bottomrule
\end{tabular}
\end{table*}

\section{Experiments}

\subsection{Experimental Setup}

We evaluate our dependency-aware code generation framework on three representative tasks that collectively assess structural correctness, semantic fidelity, and execution-level validity. These tasks cover a spectrum of code generation challenges, ranging from syntactic translation to semantic program synthesis. Following recent repository-level and structure-aware code generation studies~\cite{wang2023codet5plus,liu2024graphcoder,cheng2024draco}, we emphasize evaluation settings that require models to capture long-range interactions and preserve executable semantics rather than merely optimizing token-level similarity. The importance of robust evaluation is echoed across diverse machine learning domains, from medical image analysis~\cite{qi2025mediaug,luo2025pathohr,qi2025medconv} to computer vision~\cite{wang2026deco,zu2026end,jin2026tiny}, where benchmark design directly determines whether model improvements translate to practical utility. By including tasks that vary in complexity and ambiguity, we aim to comprehensively measure the model's capability to understand, represent, and generate executable code that adheres to both local and global dependencies. Such evaluation ensures improvements extend beyond surface-level token matching and better reflect practical software engineering utility.

\textbf{Code Translation.}
We adopt the Java$\rightarrow$C\# and C\#$\rightarrow$Java translation tasks from CodeXGLUE~\cite{lu2021codexglue}. Code translation requires models to preserve both syntactic validity and semantic equivalence across programming languages despite differences in type systems, APIs, and idiomatic usage. For instance, Java \texttt{ArrayList} operations must be correctly mapped to C\# \texttt{List} operations while preserving execution behavior. The dataset contains 10K training samples, 500 validation samples, and 1K test samples. Prior work has shown that even minor inconsistencies in variable usage, API calls, or control flow can lead to compilation or execution failures~\cite{guo2021graphcodebert,ahmad2021plbart}, highlighting the importance of modeling inter-entity dependencies and structural consistency.

\textbf{Text-to-Code Generation.}
We evaluate on the CONCODE dataset~\cite{iyer2018mapping}, where models generate Java methods from natural language descriptions. This task introduces substantial ambiguity, requiring the model to infer missing implementation details while maintaining consistency among variables, method invocations, and control-flow structures. Recent pretrained code models such as CodeT5~\cite{wang2021codet5} and CodeT5+~\cite{wang2023codet5plus} demonstrate strong generation capability on this benchmark, making it a widely adopted testbed for evaluating semantic alignment between natural language intent and executable code generation.

\textbf{Program Synthesis.}
We use the APPS benchmark~\cite{hendrycks2021apps} to evaluate execution-level correctness. APPS contains programming problems with hidden test cases, preventing direct overfitting and requiring models to generate functionally correct programs. The benchmark spans introductory exercises, interview-style questions, and competition-level algorithmic challenges, providing a realistic assessment of practical coding proficiency. Recent large code models and agentic coding systems frequently adopt APPS as a standard benchmark for evaluating executable program synthesis ability~\cite{wang2023codet5plus,zhang2024codeagent}.

\paragraph{Evaluation Metrics.}
For CodeXGLUE tasks, we report BLEU~\cite{papineni2002bleu}, Exact Match (xMatch), and CodeBLEU~\cite{ren2020codebleu}. BLEU measures token-level overlap, xMatch evaluates exact sequence replication, and CodeBLEU further incorporates syntax-aware and semantic-aware components to better assess structural correctness. For APPS, we report Test Case Accuracy and Strict Accuracy following prior program synthesis evaluation protocols~\cite{hendrycks2021apps}. These metrics jointly provide a comprehensive assessment of lexical similarity, structural alignment, and execution correctness.

\paragraph{Baselines.}
We compare against Transformer~\cite{vaswani2017attention}, CodeBERT~\cite{feng2020codebert}, GraphCodeBERT~\cite{guo2021graphcodebert}, PLBART~\cite{ahmad2021plbart}, and CodeT5~\cite{wang2021codet5}. These baselines represent widely adopted pretrained code generation architectures with varying levels of structural modeling capability, including sequence-based pretraining, data-flow-aware representations, and encoder-decoder code generation objectives. However, none explicitly enforce global inter-entity dependency consistency during generation. Comparing against these methods enables quantifying the contribution of our dependency-aware refinement mechanism.

\paragraph{Implementation Details.}
All experiments are conducted on a single server equipped with an NVIDIA A800 80GB GPU. We adopt CodeT5~\cite{wang2021codet5} as the backbone model, with the low-rank factor dimension set to the default value \(k=8\). For dependency quantization, we use \(N=8\) discrete dependency levels with values
\(
q_1, \dots, q_8 = 1, 2, 3, 4, 5, 6, 7, 8.
\)
The Jaccard consistency threshold for aligning semantic and structural dependencies is set to \(\tau = 0.8\). Unless otherwise specified, all hyperparameters are selected using validation performance and kept fixed across tasks for fair comparison.

\subsection{Main Results}

\paragraph{Code Translation.}
As shown in Table~\ref{tab:translation}, our method consistently outperforms all considered baselines across both translation directions. Similar trends have been observed in recent structure-aware code generation studies, where incorporating semantic relations and structural constraints substantially improves functional correctness beyond sequence-level matching~\cite{guo2021graphcodebert,wang2021codet5,wang2023codet5plus}. Notably, the observed gains in Exact Match (xMatch) and CodeBLEU indicate that dependency-aware modeling leads to substantial improvements in both structural correctness and semantic fidelity, extending well beyond token-level overlap. Specifically, for the Java$\rightarrow$C\# task, we achieve an increase of +1.9 BLEU and +3.4 xMatch compared to CodeT5, demonstrating that our framework significantly improves exact translation correctness. Furthermore, the improvements relative to GraphCodeBERT suggest that local data-flow modeling alone is insufficient for repository-scale or semantically complex code translation. Instead, maintaining global inter-entity consistency is essential for capturing long-range structural and functional relations across programs. These results demonstrate that explicitly modeling dependencies enables the generation of translations that are not only syntactically valid but also semantically aligned with the original source code, leading to more reliable executable outputs.
\begin{table}[h]
\centering
\caption{Evaluation results on the CONCODE text-to-code generation benchmark, where we report BLEU, Exact Match (xMatch), and CodeBLEU to measure the quality of generated Java methods from natural language descriptions.}
\label{tab:concode}
\small
\begin{tabular}{lccc}
\toprule
Model & BLEU & xMatch & CodeBLEU \\
\midrule
GPT-2 & 25.8 & 17.6 & 30.4 \\
CodeGPT & 29.4 & 18.7 & 33.2 \\
PLBART & 37.2 & 19.3 & 39.1 \\
CodeT5 & 41.2 & 22.5 & 43.8 \\
\midrule
\textbf{Ours} & \textbf{43.6} & \textbf{24.1} & \textbf{46.2} \\
\bottomrule
\end{tabular}
\end{table}

\begin{figure*}[t]
\centering
\begin{subfigure}[t]{0.32\textwidth}
    \centering
    \includegraphics[width=\linewidth,height=3.2cm]{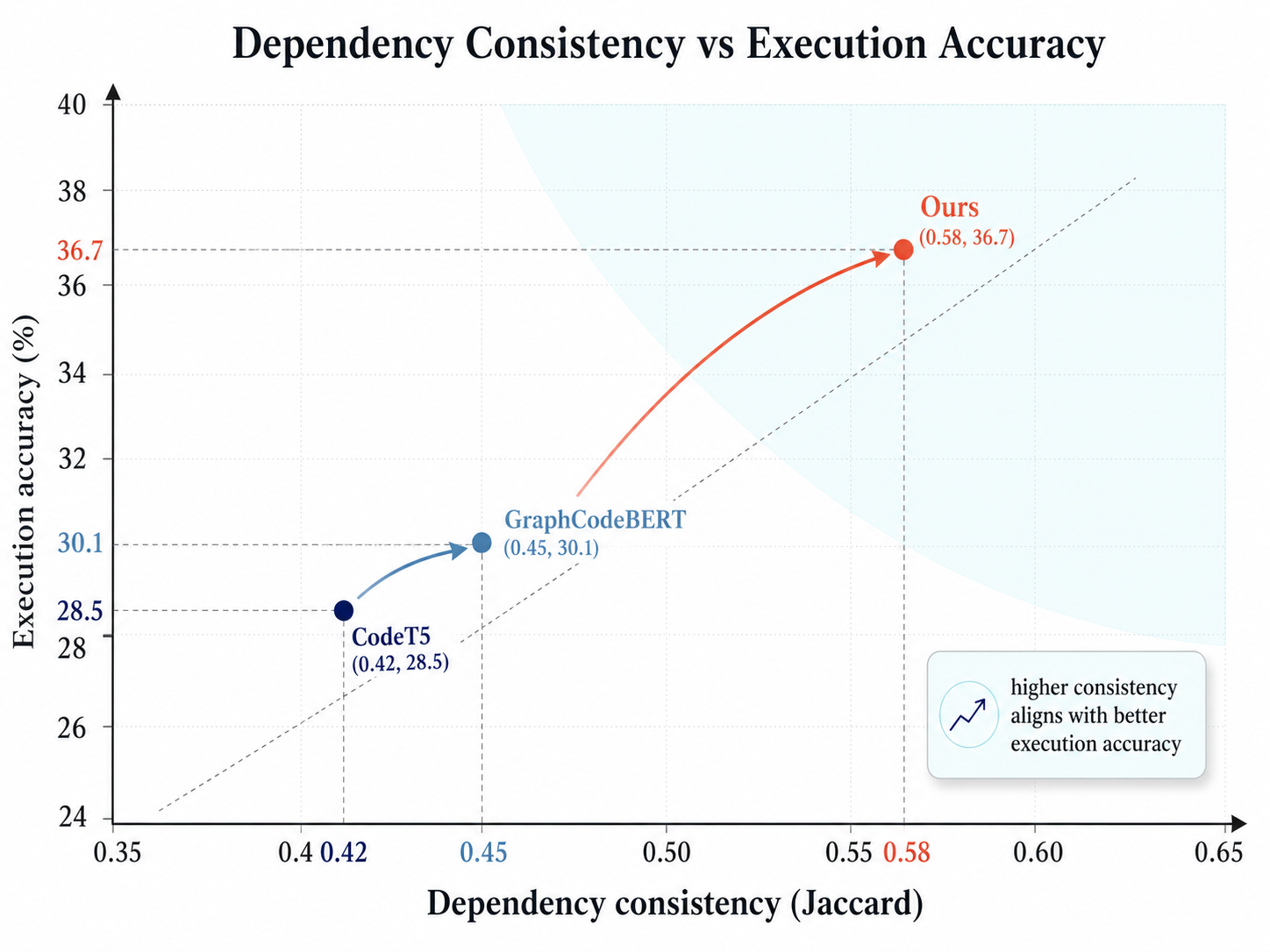}
    \caption{Dependency consistency.}
    \label{fig:dependency}
\end{subfigure}
\hfill
\begin{subfigure}[t]{0.32\textwidth}
    \centering
    \includegraphics[width=\linewidth,height=3.2cm]{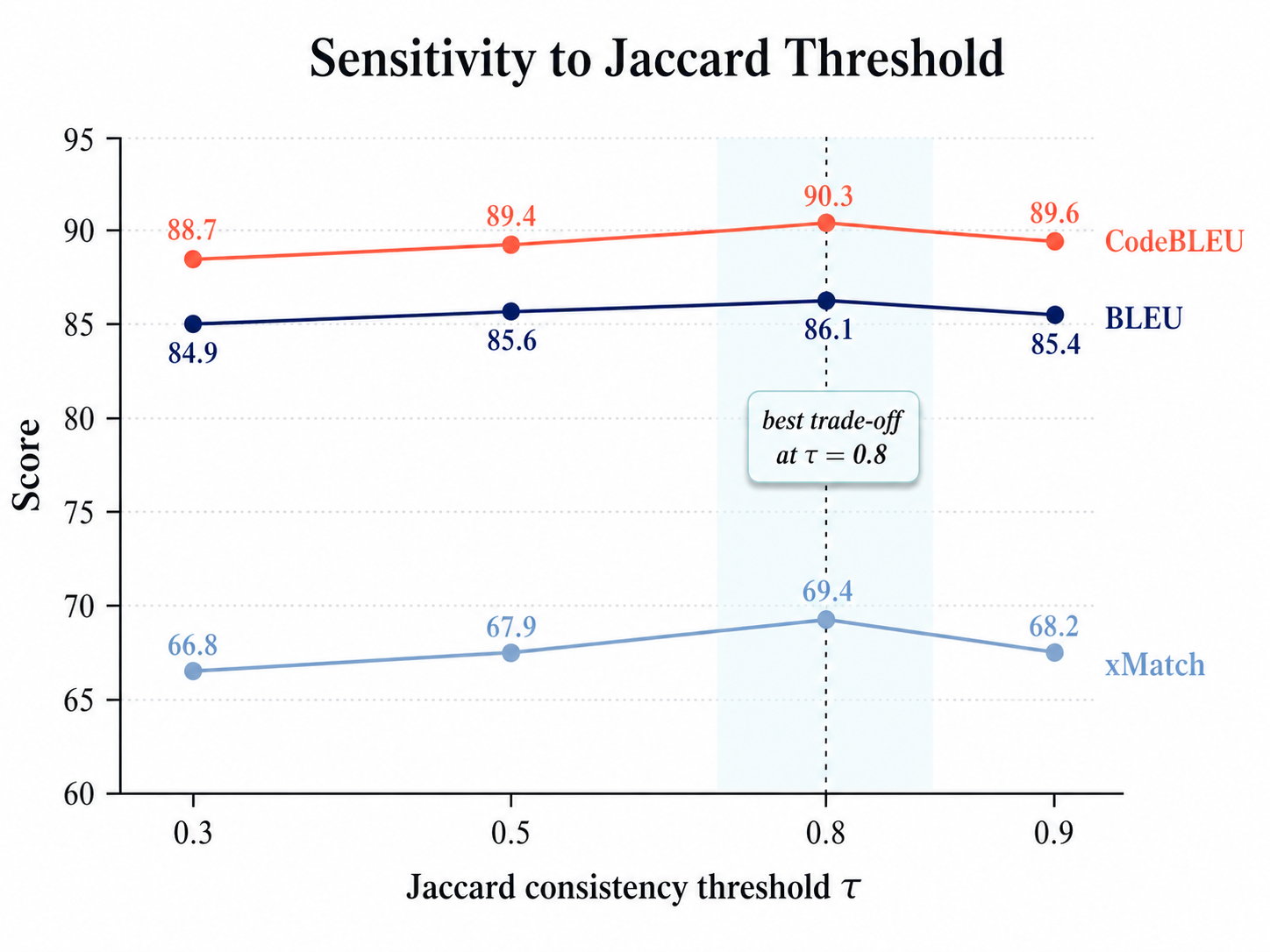}
    \caption{Sensitivity to $\tau$.}
    \label{fig:sensitivity}
\end{subfigure}
\hfill
\begin{subfigure}[t]{0.32\textwidth}
    \centering
    \includegraphics[width=\linewidth,height=3.2cm]{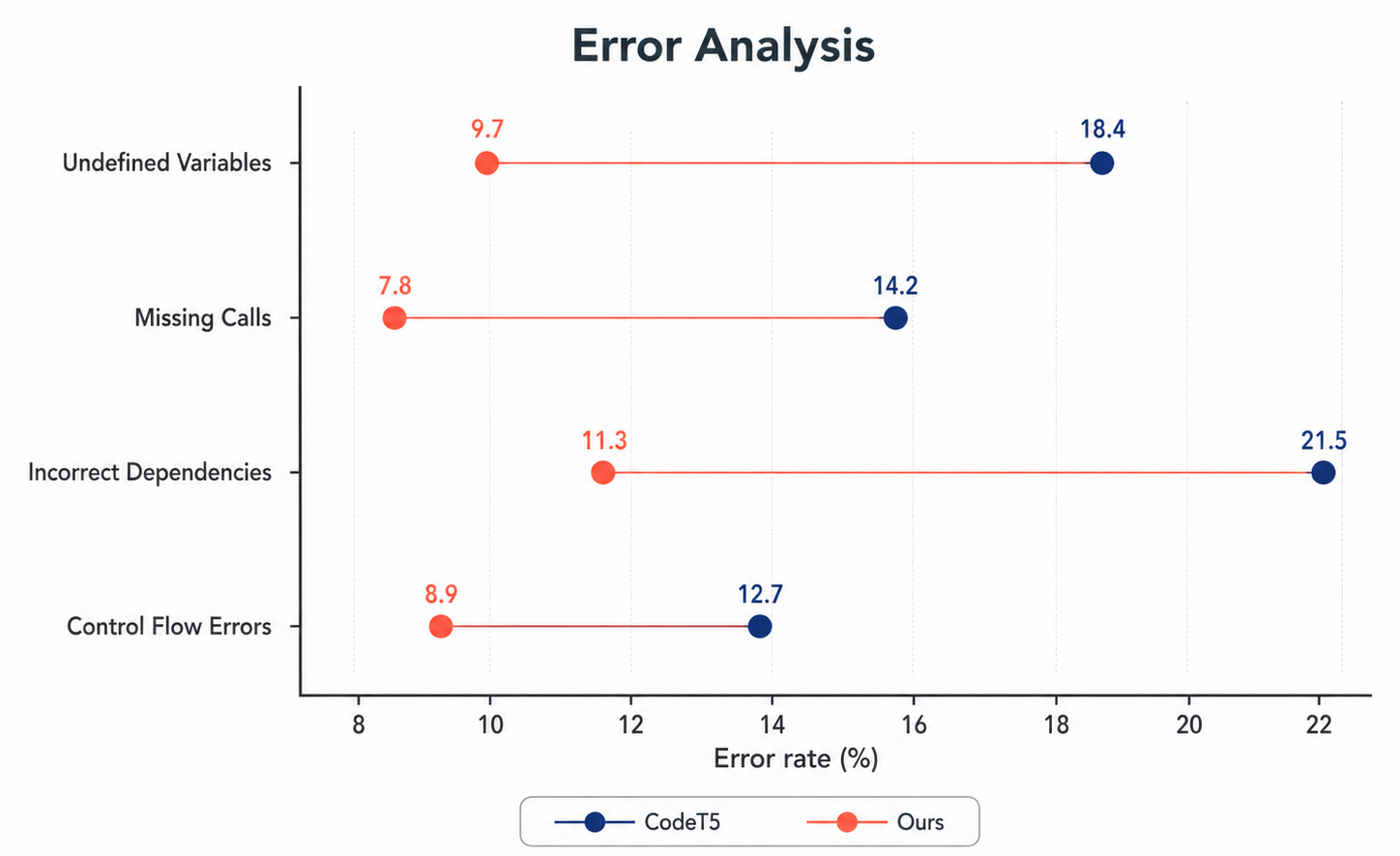}
    \caption{Error analysis.}
    \label{fig:errors}
\end{subfigure}
\caption{Comprehensive analysis of dependency-aware code generation, including dependency consistency across models, sensitivity analysis with respect to the Jaccard threshold $\tau$, and error distribution in Java$\rightarrow$C\# translation outputs.}
\label{fig:analysis_triplet}
\end{figure*}

\paragraph{Text-to-Code Generation.}

Table~\ref{tab:concode} shows that our framework achieves the best overall performance across all evaluated metrics. The improvement in CodeBLEU (+2.4 over CodeT5) indicates stronger structural alignment and more complete semantic realization, both of which are particularly important when natural language specifications contain ambiguities, underspecified behaviors, or implicit implementation details~\cite{iyer2018mapping,wang2021codet5}. Prior pretrained code models mainly optimize token prediction or local structural representations~\cite{feng2020codebert,wang2023codet5plus}, whereas our dependency-aware mechanism explicitly constrains consistency among variables, function calls, and control-flow structures during generation. As a result, the model produces code with fewer incomplete implementations, inconsistent entity usages, or logically conflicting operations. The consistent improvements across BLEU, Exact Match, and CodeBLEU further suggest that dependency-aware refinement benefits both lexical accuracy and deeper semantic coherence. These findings demonstrate that explicitly enforcing dependency consistency can effectively guide generation toward structurally reliable and semantically executable programs, improving the practical usability of generated code in real-world software engineering scenarios. Broader challenges in multimodal reasoning, including certifiable modality deletion~\cite{fu2026missing} and multimodal scale recognition~\cite{jin2026tiny}, further suggest that consistency-driven refinement principles may generalize beyond code to other structured generation tasks.

\paragraph{Program Synthesis.}

\begin{table}[h]
\centering
\caption{Results on the APPS program synthesis benchmark, reporting both Test Case Accuracy and Strict Accuracy across different difficulty levels (Intro, Interview, Competition) to evaluate execution-level correctness.}
\label{tab:apps}
\small
\resizebox{0.48\textwidth}{!}{
\begin{tabular}{lccc|ccc}
\toprule
& \multicolumn{3}{c}{Test Case Avg.} 
& \multicolumn{3}{c}{Strict Acc.} \\
Model & Intro & Inter & Comp & Intro & Inter & Comp \\
\midrule
GPT-2 (0.1B) & 5.9 & 7.2 & 4.6 & 1.1 & 0.4 & 0.0 \\
CodeT5 (0.2B) & 10.1 & 6.5 & 2.8 & 1.9 & 0.5 & 0.1 \\
\midrule
\textbf{Ours} & \textbf{11.8} & \textbf{8.3} & \textbf{3.9} 
& \textbf{3.1} & \textbf{1.1} & \textbf{0.4} \\
\bottomrule
\end{tabular}
}
\end{table}

Table~\ref{tab:apps} demonstrates that enforcing dependency consistency during generation leads to measurable gains in both partial and full execution correctness. In particular, improvements in Strict Accuracy indicate that the generated programs are not only partially correct but fully functional when executed against hidden test cases. This confirms that modeling inter-entity dependencies helps the model capture critical execution logic and control-flow relationships, thereby reducing subtle errors that could cause runtime failures. The results underscore the practical benefits of our approach: by incorporating structural constraints, the generated programs exhibit higher robustness and reliability, making the method suitable for real-world program synthesis tasks where functional correctness is paramount.

\subsection{Further Analysis}

\paragraph{Dependency Consistency.}
As shown in Figure~\ref{fig:dependency}, models with higher dependency consistency, measured by the Jaccard similarity between semantically and structurally related entities, tend to achieve higher execution accuracy. This observation is consistent with prior findings that code evaluation should account for syntactic structure and semantic data-flow rather than relying solely on surface-level token overlap~\cite{ren2020codebleu,guo2021graphcodebert}. The trend highlights the importance of maintaining coherent relationships among code entities, since programs with well-aligned dependencies are more likely to be syntactically valid and functionally correct. In particular, our approach consistently outperforms CodeT5 and GraphCodeBERT, suggesting that explicitly modeling and enforcing global entity relationships can substantially improve program quality. By capturing both strong and weak dependencies across functions, classes, and files, our framework preserves critical interactions while still allowing flexibility for optional or less central connections. These results reinforce that dependency consistency is a key factor in reliable code generation, and that approaches neglecting cross-entity relationships may produce programs with structural or functional deficiencies.

\paragraph{Sensitivity Analysis.}
As shown in Figure~\ref{fig:sensitivity}, our dependency-aware model maintains stable performance across a range of Jaccard threshold \(\tau\) values. Lower thresholds, such as \(0.3\), correspond to more relaxed dependency enforcement, which may include noisy or weakly relevant interactions and slightly reduce structural alignment. Conversely, excessively high thresholds, such as \(0.9\), impose overly strict constraints, which can over-constrain generation and limit the model's ability to produce alternative but still valid implementations. In our experiments, performance peaks around \(\tau=0.8\), indicating an effective balance between preserving critical dependencies and retaining generative flexibility. These findings show that the proposed framework is robust to moderate threshold variations and does not require fine-grained task-specific tuning. Overall, the sensitivity analysis demonstrates that dependency-aware refinement can maintain high-quality code generation under different consistency constraints.

\paragraph{Error Analysis.}
As shown in Figure~\ref{fig:errors}, our dependency-aware approach consistently reduces common code generation errors compared to CodeT5. Specifically, the rates of undefined variables, missing method calls, incorrect dependency ordering, and control-flow mistakes are substantially reduced. Since execution-based benchmarks such as APPS emphasize functional correctness through test cases rather than lexical similarity alone~\cite{hendrycks2021apps}, reducing these dependency-related errors is essential for improving practical code reliability. These improvements indicate that explicitly capturing global entity relationships helps the model maintain valid variable definitions, respect method invocation sequences, and preserve logical control flow. Overall, this analysis confirms that dependency-aware modeling directly contributes to generating more reliable, semantically consistent, and functionally correct code, thereby reducing runtime failures and improving the practical utility of generated programs.

\subsection{Ablation Study}

\begin{table}[h]
\centering
\caption{Ablation study on the Java$\rightarrow$C\# translation task, analyzing the contribution of each component in our dependency-aware framework to overall performance.}
\label{tab:ablation}
\small
\begin{tabular}{lccc}
\toprule
Model Variant & BLEU & CodeBLEU & xMatch \\
\midrule
w/o Strong (Q) & 83.5 & 88.1 & 65.3 \\
w/o Constraint & 83.2 & 87.9 & 64.8 \\
\midrule
\textbf{Full Model} & \textbf{86.1} & \textbf{90.3} & \textbf{69.4} \\
\bottomrule
\end{tabular}
\end{table}

Table~\ref{tab:ablation} shows that removing any individual component of our framework results in a noticeable decrease in overall performance, highlighting the contribution of each part to the final results. Among all components, dependency modeling proves to be the most critical, as it effectively captures the relationships between code entities and directly guides the generation process toward coherent and executable code. This observation further underscores the importance of explicitly enforcing inter-entity dependencies during code generation to maintain both structural consistency and semantic correctness. Similarly, removing other key constraints, such as decoding restrictions, can produce outputs that are inconsistent, incomplete, or less reliable. These findings demonstrate that every component of our framework plays a complementary role, and the interplay between dependency modeling, constraints, and generation mechanisms is essential for ensuring the overall reliability, correctness, and quality of the generated programs.




\section{Conclusion}
In this paper, we presented a dependency-aware code generation framework that models relationships among code entities. By representing codebases as graphs and decomposing dependencies into a quantized matrix for strong interactions and a sparse low-rank component for weaker ones, our approach captures multi-level structural relationships. We developed an efficient alternating optimization algorithm and incorporated the learned structure into generation through consistency-driven refinement. This enables generated code to better preserve semantic intent and structural constraints. The sparse triplet representation further improves storage efficiency and scalability. Experiments show gains in semantic coherence, structural consistency, and practical usability.

\section*{Limitations}
Our low-rank residual term improves robustness to incomplete or noisy dependency graphs by capturing implicit dependencies and higher-order correlations. However, static dependency extraction may still be insufficient in highly dynamic or context-sensitive settings. Incorporating runtime signals, version history, or online updates is therefore a promising direction for future work. Drawing inspiration from adaptive techniques in related domains, such as human-machine co-adaptive intent understanding~\cite{he2025enhancing}, future work could explore dynamic dependency graph updates that adapt to evolving codebases in real time.


\bibliography{custom}

\appendix

\end{document}